# Key Technology Challenges for the Study of Exoplanets and the Search for Habitable Worlds

A Whitepaper in support of the Exoplanet Science Strategy

Authors:

**Brendan Crill**, NASA Exoplanet Exploration Program, Jet Propulsion Laboratory / California Institute of Technology bcrill@jpl.nasa.gov

**Nick Siegler**, NASA Exoplanet Exploration Program, Jet Propulsion Laboratory / California Institute of Technology

**Shawn Domagal-Goldman**, NASA Goddard Space Flight Center

**Eric Mamajek**, NASA Exoplanet Exploration Program, Jet Propulsion Laboratory / California Institute of Technology

**Karl Stapelfeldt**, NASA Exoplanet Exploration Program, Jet Propulsion Laboratory / California Institute of Technology

I. Introduction

This whitepaper will outline the key technology challenges for the study of extrasolar planetary systems, and particularly for the search for life on planets in those systems. These science goals drive exoplanet missions towards capabilities to characterize exoplanets in the coming decades (See mission roadmap figure https://exoplanets.nasa.gov/internal_resources/816). These capabilities include obtaining: exoplanet spectroscopy over a broad range of wavelengths; exoplanet mass; and host star spectroscopy, particularly in the far ultra-violet. The third of these capabilities already exists; the first two have been enabled for non-habitable planets much larger than Earth. Pushing these technologies into the realm of habitable, Earth-size worlds will allow astronomers to collect data that will enable a search for life, and probe the formation histories and diversity of worlds beyond our solar system.

There are three key technology areas requiring advancement to achieve these capabilities:

1) Direct imaging of exoplanets (so as to perform reflection and emission spectroscopy as well as learn their orbital characteristics; Seager WP)
2) Transit spectroscopy (in absorption) / secondary eclipse spectroscopy (in emission; Fortney WP)
3) Stellar reflex motion (for mass measurement; Plavchan WP)

II. Technology Gaps

NASA's Exoplanet Exploration Program (ExEP) identifies technology gaps pertaining to possible exoplanet missions, works with the community to identify, track, and prioritize technology gaps, and ultimately closes the gaps via investment in technology development projects. These technologies are summarized in the ExEP's annually-updated Technology List [1] and captured in detail in their Technology Plan Appendix [2]. A possible roadmap to mature these technologies is described in [3]. The technology area performance gaps are:

**Direct imaging of exoplanets**

**Starlight suppression for reflection (or emission) spectroscopy.** Suppression of starlight in order to bring orbiting exoplanets into view requires either starlight occultation or interferometric nulling. Starlight occultation technologies include those both internal (coronagraph) and external (starshade) to the telescope. Both approaches have progressed this decade, largely due to the 2010 Astrophysics Decadal Survey's highest priority recommendation for medium-scale space activities, and the subsequent investment in technology development.

*Coronagraphs.* Ground-based telescopes with coronagraph instruments, even next generation instruments on future 30 m-class telescopes, are fundamentally limited to about $10^{-8}$ contrast sensitivities due to the residual uncorrected errors from atmospheric turbulence correction [4,5]. WFIRST's technology demonstration coronagraph, if flown, will be the first high-contrast coronagraph in space possessing wavefront-sensing and correcting optics to achieve contrast sensitivities between $10^{-8}$ and $10^{-9}$ (Bailey WP). To observe an Earth-size exoplanet orbiting in the habitable zone of a Sun-like star, however, would require sensitivities to contrast ratios of $10^{-10}$ or better (https://exoplanets.nasa.gov/internal_resources/773). This is 1-2 orders of magnitude more demanding than WFIRST's expected performance and 2 orders more than future 30 m-class ground-based telescopes.

Coronagraph performance demonstrations with apertures resembling telescopes with little or no central obscuration are already close to achieving the $10^{-10}$ contrast goal while simultaneously achieving relatively high throughput [6]. To enable the performance of next-generation coronagraphs to reach this goal, an ExEP-led facility called the Decadal Survey Testbed is being commissioned and is expected to have first results in CY 2018.

Future large space telescopes may be composed of segmented mirrors and secondary mirror obscurations. A number of efforts are attempting to address these additional challenges in achieving the same contrast goals while maintaining reasonable throughput and robustness to wavefront errors. In 2016, the ExEP chartered the Segmented Coronagraph Design & Analysis study to work with leading coronagraph designers. At the time of this writing, there is one candidate that is meeting the requirements [7]. Modeling results are due this summer. If successful, the masks and optics for successful designs will be fabricated and tested in air and then tested in the ExEP Decadal Survey Testbed under vacuum in 2019. Before the 2020 Decadal Survey we should know how well coronagraphs will work with mirrors, whether unobscured or obscured, monolithic or segmented, for future exoplanet missions.

*Starshades.*  The starshade is currently being advanced under a single ExEP technology development activity whose objective is to advance five key technologies to TRL 5 (Ziemer WP). WFIRST is being used as a reference mission for the design and engineering work (a starshade, however, is not baselined for WFIRST). While a starshade's optical performance can never be demonstrated at full scale on the ground, a preliminary assessment [8] has developed design models with error budgets predicting better than $10^{-10}$ contrast (or $10^{-9}$ when expressed in terms of starlight suppression; see [2] for definitions). A sub-scale validation demonstration has already achieved $4.6 \times 10^{-8}$ starlight suppression at flight Fresnel numbers and is expected in CY18 to demonstrate the $10^{-9}$ starlight suppression goal. However, to test at these regimes and operate within a practically-sized testbed, the demonstration is being conducted with only a 25 mm starshade (testbed is already 77 m long; testing large sizes require very long testbeds as separation between the starshade and "telescope" increase by the square of the starshade radius). Hence, confidence in these sub-scale starshade demonstrations to represent full-scale performance will depend on their ability to validate their performance models. Despite diffraction theory predicting optical performance to be independent of scale, additional suppression demonstrations are planned to be completed by CY20 at different wavelengths, starshade sizes, and a range of key perturbations to demonstrate the robustness of the models.

Another key technology being advanced targets reducing the scattering of sunlight off the starshade's petal edges. Materials that are sufficiently thin, low-reflectivity, and suitable for stowage are being investigated as "optical edges". Amorphous metals are a promising candidate and are currently being tested. Unlike other large structural deployments, the starshade requires precise and stable positioning of a 30 m structure to better than 1 mm. A half-scale or larger prototype is planned to be demonstrated to meet deployment tolerances

**Contrast stability.**  Due to the extremely low rate of photons detected from distant exoplanets (in the range of about a photon per minute(s) in the case of the WFIRST coronagraph), performing spectroscopy at a sufficient signal-to-noise ratio will require the contrast to be maintained for long integration periods. In the case of coronagraphy, this is expected to translate to sensing and controlling wavefront errors typically between 10-100 pm rms for a telescope and instrument system [9]. While instrument-level lab demonstrations to date are within factor of a few of this requirement, this is 1-2 orders of magnitude more demanding than the performance of current and upcoming space telescopes.

This level of extreme wavefront stability must be maintained as the space observatory and its coronagraph experience typical environmental disturbances during operation - dynamic jitter and thermal drifts. Large mirrors, both monolithic and segmented, will be challenged by the need to achieve a stable back-structure and segmented ones will need to maintain a large number of individual segments as a single paraboloid. Due to these tight stability requirements, coronagraphs can no longer be designed as separate payload instruments but rather along with the observatory as a single system. On-going analyses by the HabEx and LUVOIR study design teams are determining the best approach to these challenges for space-based telescopes of a range of sizes. Their work, and the assessment thereof, will determine the likelihood of these telescope systems meeting the very demanding wavefront error stability requirements.

In the case of a starshade-only mission, telescope stability requirements are significantly looser and do not exceed the state-of-the-art (SOA). Solutions for sensing and alignment control between the two spacecrafts have been developed and subscale demonstrations are being conducted in the lab. Thus, the technology development for missions that utilize starshades falls primarily on the starshade itself, and not on the optical telescope assembly.

**Detection sensitivity.** The low flux from the rocky exoplanets requires a detector with read noise and spurious photon count rate as close to zero as possible, and that maintains adequate performance in the space environment. The SOA is dependent on the wavelength band but detectors must perform at or near the photon counting limit from the UV through the NIR for current mission concepts. Across this wavelength range, the SOA detectors are semiconductor-based devices. WFIRST's electron multiplying charge coupled device (EMCCD) detectors have achieved adequate noise performance in the visible band, though longer lifetime in the space radiation environment is needed. Similar EMCCD devices, with delta doping, may already have adequate performance in the near-UV. HgCdTe detectors are the SOA in the NIR. JWST/MIRI's detectors are expected to establish the SOA in MIR detection sensitivity, and future direct imaging is likely to require detectors that exceed it. It is likely that the detection sensitivity gap can be closed in the next decade, as a range of choices are close to meeting the requirements (Vashist WP)

**Angular resolution and collecting area**. Large space telescopes offer many benefits in the determination of exoplanet habitability such as tighter point spread functions (greater sensitivity to faint objects), improved spatial resolution (to probe the habitable zones of more distant stars), improved spectral resolution (for better feature signal-to-noise), shorter integration times (offering the possibility of studying different faces of a rotating planet in a nearby star system), and better rejection of the extended diffuse brightness of exozodiacal light (that could obscure exoplanets). All else being equal, these advantages allow larger-aperture telescopes to obtain a larger exoplanet yield [10], a benefit that may prove to be very important if the frequency of habitable planets is small. They also allow larger-aperture telescopes (again all else being equal) to obtain better data for a given target.

The largest monoliths flown in space are the 2.4 m Hubble Space Telescope, optimized for visible and UV astronomy, and Herschel's 3.5 m telescope, optimized for the far-IR. The James Webb Space Telescope will establish the SOA in mid-IR space telescopes with a 6.5 m primary mirror made up of 18 co-phased hexagonal beryllium segments. Current large mission concept studies range from 4 m monoliths to 15 m segmented telescopes.

Large glass monoliths are commonly fabricated for ground-based telescopes. If future heavy-lift launch vehicles like the Space Launch System become a reality then the opportunity for a 4- to 8- m-class monolith becomes a possibility. Large monoliths will advance exoplanet science but will not directly lead to subsequent larger telescope architectures. Mirrors constructed from one-meter class silicon carbide or glass segments have fabrication heritage and appear to be promising options if the design teams can show there is sufficient control authority to meet the contrast goals.

**Transit/secondary eclipse spectroscopy**

**Spectroscopic Sensitivity.** To enable precise transit or secondary eclipse spectroscopy of exoplanets, the detector response must exhibit photometric stability over the time scales of a transit, typically hours. Spitzer/IRAC has achieved photometric stability of order 60 parts per million on transit time scales. JWST/MIRI is expected to achieve stability between 10-100 ppm. A stability of 5-10 ppm in the mid-IR is needed in order to measure the atmospheres of Earth-sized planets transiting nearby M-dwarfs (Fortney WP).

The path to close the technology gap in transit spectroscopy of Earth-sized planets is a challenging one. First, astrophysical limits should be examined further to find the likely fundamental limits to stellar stability. The sources of instability in detector/telescope systems must be studied to determine where future technology investments will be most effective. Photometric instabilities of a mid-IR detector system may be driven by fundamental detector materials properties, cryogenic detector readout circuitry, or other instabilities in the system. This should be investigated along with modeling the on-orbit calibration, which will mitigate the detector requirements to some level. Valuable lessons will be learned from performing these measurements with JWST/MIRI in the early 2020s (Vashist WP).

**Stellar reflex motion**

**Radial stellar motion sensitivity.** Radial velocity (RV) measurements of the reflex motion of a star is a way to infer the minimum mass and orbital parameters of planets orbiting the star. The reflex motion of a Solar-mass star due to an orbiting Earth-mass planet at 1 AU is ~10 cm/s over 1 year, and both measurement and systematic errors must be kept below that. The HARPS instrument has recently detected 40 cm/s signals [11]. The next generation of ground-based RV instruments coming online in the next two years are expected to achieve 20-30 cm/s instrumental sensitivity per measurement.

The biggest uncertainty in closing this gap is understanding the astrophysical limits due to natural stellar jitter. At this point the path forward to achieving 1 cm/s sensitivity and closing the gap is unclear. The challenge is likely to be mitigated with a broad wavelength range of observations and observing from space, and may be better understood upon completion of a NASA-chartered probe study, and through experience at mitigating systematics errors in ground-based RV instruments measurements (Plavchan WP).

**Tangential Stellar Motion Sensitivity**. By performing sensitive astrometry of a star over time, the mass and orbital parameters of orbiting exoplanets can be measured. GAIA's initial data release achieved typically 300 microarcsecond ($\mu$as) position error, but subsequent data releases are expected to achieve 10 $\mu$as sensitivity in the positions of many stars, enough to reveal many Jupiter-mass exoplanets. A precision of 0.3 $\mu$as per measurement is needed in order to enable the detection of Earth-mass planets at a distance of 10 pc.

It is possible that astrophysical limits due to variable stellar surface structure may prevent astronomers from reaching this precision. The inherent instabilities of stars needs further understanding and sources of instrument instability and the ability to calibrate them using techniques such as interferences fringes or diffractive pupils should be modeled (Bendek WP).

## III.    Future Technology Needs

Any of the large exoplanet missions under study leading up to the 2020 Decadal Survey (e.g., LUVOIR, HabEx, OST), whose technology needs we outlined above, will be sensitive to detecting biologically-produced gases in the atmospheres of Earth-size planets in the habitable zone of their stars. All of these missions have a strategy to rule out known abiotic mechanisms for producing those gases. However, the history of biosignature claims inevitably leads to the discovery of new abiotic means to produce data originally claimed to be a biosignature [12]. The resolution of these claims may require a subsequent mission to confirm or rule-out life before the question "Are we alone?" is finally answered.

If the first signs of life are detected via features that exist in the UV-NIR (e.g., via HabEx or LUVOIR), a subsequent mission could confirm those biosignatures in the mid-IR. This would allow a more thorough search for methane ($CH_4$), and carbon dioxide ($CO_2$). Because it becomes impractical with a single aperture telescope to resolve the habitable zones of nearby stars at 15 $\mu$m (where the important $CO_2$ feature is present), interferometry may be the approach of choice where the resolution is set by the baseline between multiple apertures rather than the size of the individual telescopes. Interferometry was studied in the early 2000's as part of the Terrestrial Planet Finder Interferometer concept, whose study identified technology gaps in path length stability, detector sensitivity, passive and active cryogenic cooling, and formation flying. Investments in technologies to close these gaps should begin no less than 15 yr before mission start. Interferometry is also a key technology for the "Visionary Era" of NASA's Astrophysics Roadmap [13], including an exo-Earth mapper that would achieve spatial resolution across the surface of an Earth-sized exoplanet.

Alternatively, a search for secondary biosignatures could occur with a more sensitive telescope that allows for the detection of smaller spectral features. This could be achieved by larger single-aperture optical telescopes. Eventually, such telescopes may exceed the largest possible launch fairing (or the autonomous deployment risk may be unacceptable). This would create a compelling need for in-space assembly. The point at which in-space telescope assembly becomes more affordable or reduces risk to a level more acceptable than autonomous deployment has not yet been determined (Mukherjee WP). A study to address this trade would be informative for future space telescopes.